# Physical origin of shear-banding in jammed systems

P. Coussot, G. Ovarlez

Université Paris-Est, Laboratoire Navier (UMR 8205 LCPC-ENPC-CNRS), Champ sur Marne, France

**Abstract**: Jammed systems all have a yield stress. Among these materials some have been shown to shear-band but it is as yet unclear why some materials develop shear-band and some others do not. In order to rationalize existing data concerning the flow characteristics of jammed systems and in particular understand the physical origin of such a difference we propose a simple approach for describing the steady flow behaviour of yield stress fluids, which retains only basic physical ingredients. Within this frame we show that in the liquid regime the behaviour of jammed systems turns from that of a simple yield stress fluid (exhibiting homogeneous flows) to a shear-banding material when the ratio of a characteristic relaxation time of the system to a restructuring time becomes smaller than 1, thus suggesting a possible physical origin of these trends.

## 1. Introduction

A wide range of materials (concentrated emulsions, foams, colloids, etc) are considered as jammed systems [1] but they have very different types of structures at a local scale so that the unity of these systems mainly relies on their (macroscopic) mechanical properties. Their basic, common, rheological property, directly related to their jammed character, is their yield stress, namely the fact that they can flow only when submitted to a stress larger than a critical value. With the development of techniques providing information about the local flow characteristics (within the sample), it was shown that various such systems can develop shear-banding (coexistence of sheared and unsheared regions) at sufficiently low apparent shear rates even in homogeneous stress field [2]. This effect, which strongly contrasts with the usual assumption in (apparent) rheology of complex fluids, finds its origin in that the material is unable to flow steadily at a shear rate smaller than a critical, finite value [3]. As yet, although this might be of major importance for our understanding of these systems it remains unclear why some materials exhibit shear-banding and other materials don't exhibit shear-banding. Is this difference due to some specificity of the interactions at a local scale?

Recently it was suggested that attractive (jammed) systems would tend to develop shear-banding whereas repulsive systems would not [4-5], but no physical explanation of the process leading to such trends was suggested. Besides, different types of modelling approaches [6-8] were able to predict the possibility of shear-banding for an appropriate set of parameters in purely phenomenological model [6], or mesoscopic approaches, i.e. SGR model [9] or Fluidity model [10]. Their predictions rely on the coupling of an equation providing the stress with an equation describing the evolution of a kind of state variable with additional parameters which cannot easily be given clear physical meaning. The critical point is that, according to these models, it seems necessary to introduce a specific ingredient, via a new parameter, in the model, in order to get shear-banding. As a consequence these models do not seem to provide a unity of shear-banding and non shear-banding materials within a single physical scheme.

Here we propose an approach which, rather than attempting to describe in detail the different aspects of the behavior of specific material types, aims at providing some physical explanation of the above trends, namely the origin of homogeneous flow or shear-banding, and how the intensity of the shear-banding effect varies with some physical properties of the material. This approach relies on a simple frame which retains only basic physical ingredients in such a system. Our model predicts that in the liquid regime the behaviour of jammed systems turns from that of a simple yield stress fluid exhibiting homogeneous flows to a shear-banding material when the ratio of a characteristic relaxation time of the system to its restructuring time becomes smaller than 1. We show that this result is qualitatively in agreement with existing data. We emphasize that our model is not aimed at providing a new general theory for describing the detailed behavior of these systems; our goal is to get a physical explanation why some materials exhibit shear-banding while others do not. The oversimplifications are just the price to pay to encompass jammed systems of very different structures.

## 2. A simple sketch of jammed systems

2.1 Basic assumptions

Jammed systems include a wide range of materials with very different structures. Some of them like concentrated emulsions or foams are made of soft elements squeezed against each other; some others, such as some colloidal suspensions, are made of solid particles able to form attractive links with other particles. For these various materials the origin of jamming at a local scale differs, but in all cases each jammed element is in a potential well resulting from the repulsive or attractive forces of its neighbours, and it unjams when it moves out of this potential well. From this point of view we have a unity of these systems. In any case the potential well tends to *attract* the element when it is in an appropriate position, and in contrast this specific well has no more action on the element when this element is sufficiently "far away" from this potential well. In fact for some systems (concentrated emulsions, foams) the time during which an element is out of any potential well is likely very short but it is still finite. Thus whatever the origin of jamming a jammed system may be seen as made of elements with attractive links able to break and reform with specific characteristics depending on each system.

The next step is a mean field approach in which we assume that the mechanical behavior is given from the average of the behavior of elementary volumes typically containing a pair of elements (that will be referred to as particles in the following) immersed in the liquid (of viscosity $\mu_0$). Thus we leave apart collective effects, which no doubt have a quantitative impact on some physical properties of most jammed systems but which will appear not critical in our context. We consider that in a given elementary volume the particles can be in two different states: linked or unlinked. In the first configuration (linked particles) the stress needed to shear the volume at a rate $\dot{\gamma}$ is the sum of an elastic stress inducing a deformation of the particle pair (which has a shear modulus $G$) and a stress needed to shear the liquid. The link breaks when the elastic stress reaches a critical value $\tau_c = G\gamma_c$ associated with a critical deformation $\gamma_c$. When the link is broken there is no more interaction between the particles so that the stress needed to maintain flow is only a simple viscous one. For the sake of simplicity we assume that the stress needed to shear the liquid fraction in the linked and unlinked state is identical, since in both cases we are dealing with a concentrated suspension of particles in a simple liquid: this stress writes $\mu\dot{\gamma}$, in which $\mu$ is proportional to $\mu_0$.

In the initial state, i.e. after some time at rest, the particles are linked. When a stress $\tau$ lower than $\tau_c$ is applied at the initial time the material response is that of a viscoelastic solid for which in steady state (more precisely for $t \gg \mu/G$) we have $\dot{\gamma} = 0$. For a stress larger than $\tau_c$ there is some viscoelastic behaviour in the very first time (with a characteristic time $\mu/G$) then in steady state the link is broken and we have $\dot{\gamma} = \tau/\mu$. The apparent flow curve of the material obtained from an increasing stress ramp thus resembles the typical flow curve of a yield stress fluid. Note however that the transition to the liquid regime is here rather abrupt: just beyond the yield stress the material flows at a finite shear rate $\dot{\gamma} = \tau_c/\mu$. This abrupt liquefaction is analogous to what is observed for various thixotropic colloidal systems [3].

There nevertheless remains a critical problem due to the fact that the above liquefaction process is not reversible. As soon as the link has been broken the system is and remains Newtonian: if the applied stress is then decreased towards zero the system goes on flowing steadily down to infinitely small shear rates, there is no more yield stress. This means that a critical physical ingredient is not taken into account here: the material in a liquid state should be able to reversibly turn to a solid under some conditions, the link should be in some way restorable. Various physical processes may lead to link restoration with very different kinetics: for example progressive aggregation in colloids [11], structure rearrangements in foams [12-13], etc. Here we use a very simplified, generic description of these phenomena: since the particles are on average always close to each other in the elementary volumes the average (in space) particle distribution is constant whatever the flow history, and a broken link is restored after a characteristic time $\theta$. This way we model very different restructuring processes at a local scale with a single, simple approach. This strong simplification is justified with regards to our aim to understand why, among the wide range of soft-jammed systems, some exhibit shear-banding while others do not. Besides, the independence of $\theta$ on the flow history is justified by the fact that restructuring is essentially a matter of spatial configuration, i.e. it essentially occurs when the particles are in appropriate relative positions. Obviously flow intensity plays some role but in our model it is indirectly accounted for by its impact on the relative times spent in linked (restructured) or unlinked configurations.

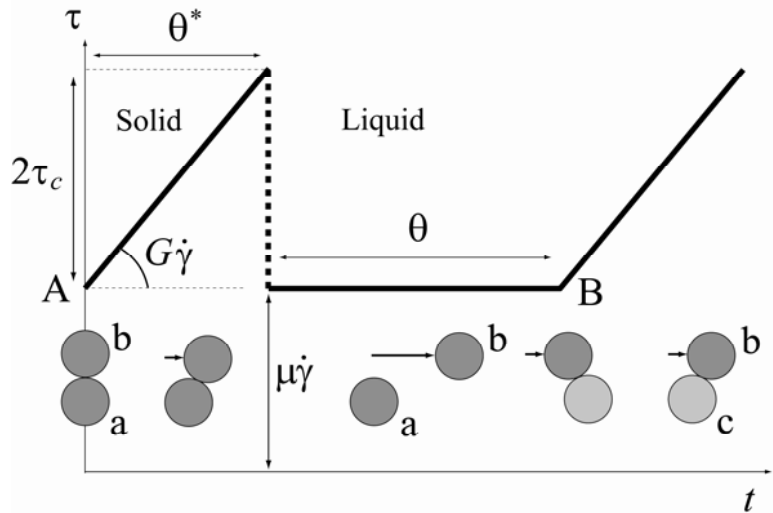

**Figure 1**: Response of the model under imposed shear rate: elementary steps of structure and stress evolution at a local scale.

2.2 Flow curve predicted by the model

Let us now detail the predictions of the model in the liquid regime under controlled shear rate.

The typical path followed by a given element is illustrated in Figure 1: elastic energy storage in the link over a time $\theta^* = \gamma_c/\dot{\gamma}$, then breakage, then link restoration after a time $\theta$. Note that now we write the critical stress for breakage $2\tau_c$ instead of $\tau_c$ to simplify the subsequent presentation. An important point is that due to the disorder of the system we assume that on average all the configurations of elements exist at any time. As a consequence, at any time in the liquid regime the stress levels to be applied on the elements are distributed uniformly over the range from A to B (see Figure 1), and the macroscopic stress due to particle link is $\tau_c \theta^*/(\theta + \theta^*)$. Actually this scheme has some analogy with the approach developed for foams initially by Princen [12]. In that case, at the bubble scale the stress is progressively increased then abruptly drops to a low value after a T1 event, but averaging over a large number of events at different stages provides a constant finite stress. In our model this situation corresponds to $\theta << \theta^*$. Also note that due to the linearity of the problem a similar result for the above average stress would be obtained for a distribution of elastic moduli with an average value $G$ (assuming a constant value for $\gamma_c$).

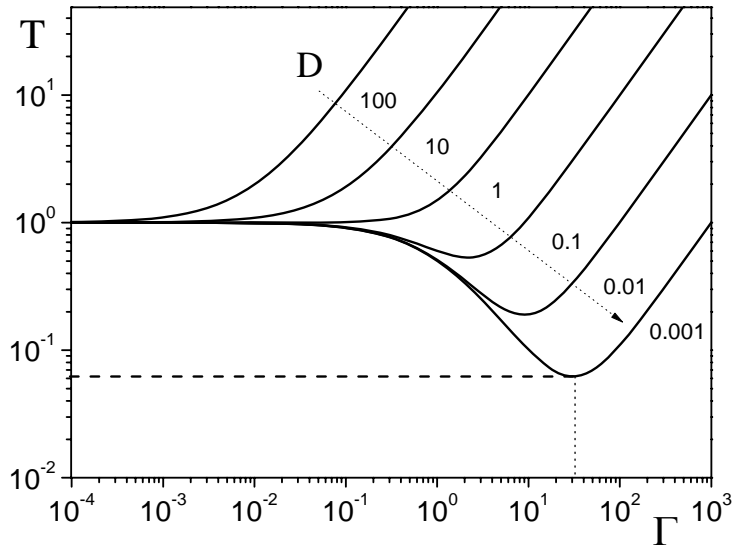

**Figure 2**: Dimensionless flow curves predicted by the model (equation (1)) for different values of $D$.

The total shear stress over the sample volume in the liquid regime is finally:

$$\tau_{mean} = \mu\dot{\gamma} + \frac{\tau_c}{1 + \dot{\gamma}\theta/\gamma_c} \qquad (1)$$

In dimensionless form equation (1) writes $T = D\Gamma + 1/(1+\Gamma)$, with $T = \tau/\tau_c$, $\Gamma = \dot{\gamma}\theta/\gamma_c$ and $D = \mu/G\theta$. This model leads to two types of behaviour (see Figure 2) depending on the value of the dimensionless number $D$: when $D > 1$ the shear stress vs shear rate curve monotonously increases; the material is apparently (from this simple description of the liquid regime) a *simple yield stress fluid*, with a yield stress equal to $\tau_c$; when $D < 1$ the shear stress vs shear rate decreases for $\dot{\gamma} < \dot{\gamma}_c = (\gamma_c/\theta)\left[1/\sqrt{D} - 1\right]$ then it increases; the *flow curve has thus a minimum* for $\dot{\gamma}_c$. Nevertheless it is a well-known result in rheology [14] that the decreasing part of a flow curve cannot correspond to the intrinsic behaviour of the material, some instability necessarily occurs, typically in the form of shear-banding. Indeed a possible, stable solution to the flow problem is a partial shear over a total thickness $h$ at the critical shear rate (i.e. in the increasing part of the flow curve) and no shear elsewhere, with the so-called *lever rule*: $h = H\dot{\gamma}_{app}/\dot{\gamma}_c$, in which $H$ is the sheared gap and $\dot{\gamma}_{app}$ the apparent shear rate (ratio of the relative velocity of the tools to the gap). Such a solution has the interest to minimize the viscous dissipation and it was effectively observed in several cases [2-3] although existing theories do not necessarily predict such a distribution [15].

Actually most jammed systems exhibit a shear-thinning behavior even at shear stresses much larger than the yield stress. In order to provide a more precise expression for a quantitative comparison with some data (see below) the model may be adapted by using a viscous term of the form $k\dot{\gamma}^n$. The introduction of this non-linear term in the constitutive equation has no immediate explanation but there exist some approaches suggesting that it would find its origin in some evolution of the flowing structure with shear rate [16]. In that case the same qualitative predictions as above are obtained with now $D = \mu/G\theta$, with the apparent viscosity $\mu = nk(\gamma_c/\theta)^{n-1}$. The critical value for the transition is $D_c = 0.25(1+n)^{1+n}(1-n)^{1-n}$, which ranges from 0.25 to 1 when $n$ ranges from 0 to 1.

2.3 Flow regimes: transition from shear-banding to no shear-banding

We are now able to predict the conditions under which shear-banding occurs in the context of this simple approach: it requires that at the same time $\mu/G < \theta$ and $\dot{\gamma}_{app} < \dot{\gamma}_c$, which defines a specific region in a ($\mu/G$, $\theta$, $\dot{\gamma}_{app}$) diagram. As the system approaches the boundaries of this shear-banding region, either by increasing $\mu/G$, decreasing $\theta$ or increasing $\dot{\gamma}_{app}$, the thickness of the sheared band increases and finally the flow becomes homogeneous. These results may be represented in a two-dimensional phase diagram (see Figure 3).

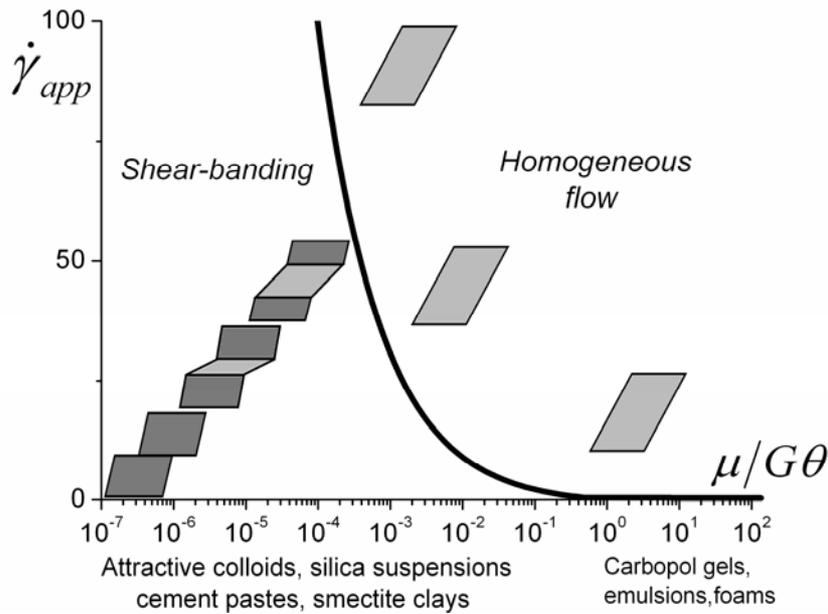

**Figure 3**: Distribution of shear-banding and homogeneous regimes as a function of the apparent (macroscopic) shear rate and the ratio of the two characteristic times of the material according to the model (assuming $\gamma_c = 1$).

At this stage we have obtained two important results:

- A simple model with attractive links only is able to reproduce the different flow types (shear-banding or homogeneous flows);

- The transition from homogeneous flows to shear-banding is governed by a parameter $D$ which is the ratio of two characteristics times of the material: the "restructuring time" $\theta$ which relates to the kinetics of the particle link restoration after breakage, and the "relaxation time" $\mu/G$ related to the viscoelastic behavior of the material.

## 3. Comparison with experimental data

Due to the specific structure of concentrated emulsions, foams or microgels the relaxation time and the restructuring time of these materials have a similar physical origin related to the viscoelastic characteristic time $\mu/G$: a recovery of some deformation of an element within the surrounding liquid and under the action of the neighbouring elements acting on it. As a consequence we can expect some similar values of these two characteristic times for various such materials, leading to a value for $D$ of the order of 1. This is indeed what we find from literature data. The restructuring time $\theta$ for such systems can be estimated from measurements made [17] for foams, of the characteristic times for T1 events associated with an elementary unjamming event, which typically takes values below 1s. Besides, for concentrated emulsions, microgels or foams we have $k/\tau_c$ of the order of 1 (in proper units). As a consequence, for these materials $D$ is larger or very close to $D_c$ so that either there is no critical shear rate or it is so small that under usual flow conditions $\dot{\gamma} >> \dot{\gamma}_c$ and there is no shear-banding. Such results are consistent with the NMR observations of the effective flow curve (determined from local velocity measurements) for a "pure" concentrated emulsion [18], a Carbopol gel [5] and various home-made foams [19] (previous data on a specific foam [20] were probably affected by experimental artefacts).

We can act on the restructuring process of the pure emulsion by adding colloidal particles which tend to create attractive links between the droplets [18]. There is now some significant restructuring process occurring in the "loaded" emulsion as evidenced by the significant increase of the elastic modulus over duration of several hundreds of seconds. Thus here $\theta$ is much larger than say 10s, and the parameter $\tau_c$ related to the strength of the link is also larger than for the pure emulsion so that we get $D < 0.025$, which predicts a strong shear-banding effect in steady state. This result is in agreement with experimental data [18]: the effective flow curve of the loaded emulsion starts beyond a critical shear rate but it is similar to that of the pure emulsion at sufficiently large shear rates, when the additional restructuring process becomes negligible. Note that these data were obtained from direct MRI observations of the local flow characteristics which provide the effective flow curve of the material (shear stress vs shear rate at different places within the material), and not the apparent (macroscopic) one which may suffer from interpretation bias. Finally our model can quantitatively predict this transition from a simple yield stress fluid to a shear-banding material by an appropriate increase of the values for $\tau_c$ and $\theta$ (see Figure 4).

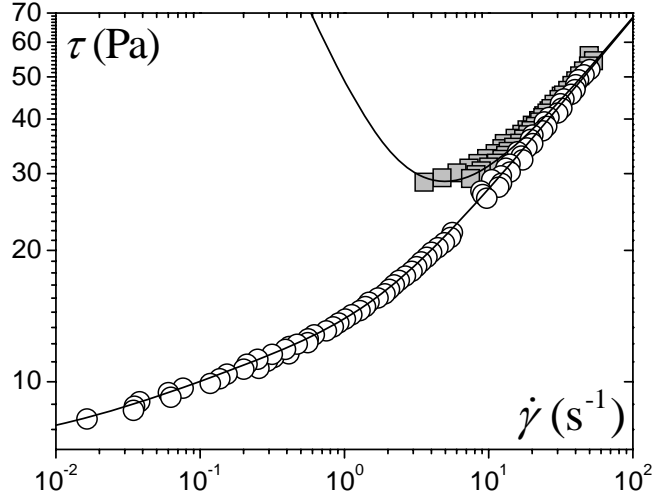

**Figure 4**: Steady-state shear stress vs shear rate for the pure (circles) and loaded (grey squares) emulsions of [18]. The solid lines correspond to the model fitted to data keeping the same viscous term ( $k = 10.8 \text{Pa.s}^{0.4}$, $n = 0.4$ ) and $\tau_c = 6.3 \text{Pa}$ ; $\theta/\gamma_c = 1\text{s}$ for the pure emulsion and $\tau_c = 420 \text{Pa}$ ; $\theta/\gamma_c = 10\text{s}$ for the loaded one.

Besides we have various colloidal suspensions which, as a first approximate, exhibit a similar apparent behavior in steady state flow. The strong difference once again lies in the restructuring process for example evidenced by the evolutions of the material properties at rest. Typically for various materials such as mustard, cement paste, clay suspensions, drilling fluids, silica suspensions, the characteristic times of restructuring at rest are of the order or larger than 10 or 100s, leading to values $D \ll D_c$. In that case the model predicts the existence of a critical shear rate for these materials, as observed experimentally [2]. In one case a restructuring process was only observed during flow, leading again to a shear-banding effect [8], and using the corresponding characteristic time one also finds $D \ll D_c$.

We can test further the ability of the model to predict the shape of the flow curve of typical shear-banding systems. In that aim we carried out tests with bentonite suspensions (see material characteristics in [21]) at different solid volume fractions (between 3 and 7%) and determined their effective ("local") flow curve in steady state from MRI measurements with the equipment and technique described elsewhere [21]. The material was initially presheared at a value such that it was fully sheared within the gap of the Couette rheometer. Then different lower rotation velocities of the inner cylinder were imposed, and the corresponding velocity profiles in steady state, reached after about 30 min. of flow, were measured. This procedure ensures that the measured flow curves correspond to the steady state liquid regime as described by our mean field approach. The validity of the results for each solid fraction was confirmed by the consistency of the effective flow curve data obtained under different rotation velocities of the inner cylinder. The flow curves exhibit a critical shear rate which increases with the solid fraction (see inset of Figure 5). For these materials the elastic modulus at rest significantly increases in time with characteristic times much larger than 10s [21], and $D$ is likely much smaller than 1. In that case there is a minimum in the flow curve at $\dot{\gamma}_c = (\gamma_c/\theta)D^{-1/1+n}$, which is much larger than 1, and the second term of (1) is well approximated by $\tau_c \gamma_c / \dot{\gamma} \theta \gg 1$. Then, by rescaling $\dot{\gamma}$ and $\tau$ respectively by the simply measurable quantities $\dot{\gamma}_c$ and the minimum stress value in the flow curve, $\tau_{c,\min}$, the constitutive equation now writes $(1+1/n)\Gamma \approx \Gamma^n/n + \Gamma^{-1}$, with here $\mathrm{T}=\tau/\tau_{c,\min}$ and $\Gamma = \dot{\gamma}/\dot{\gamma}_c$. In that way we indeed obtain a master curve with our experimental data, which means that $n$ is independent of the solid fraction (see Figure 5), and this curve can be very well fitted by the model by choosing an appropriate value for $n$.

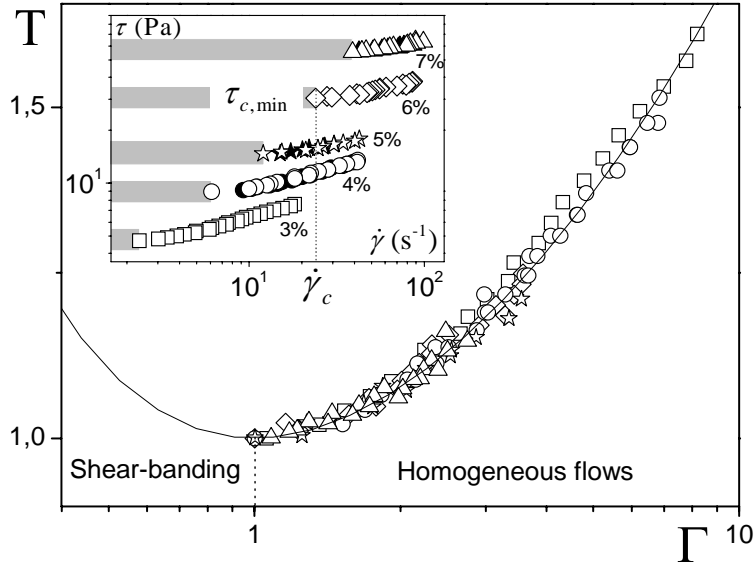

**Figure 5**: Flow curves for bentonite suspensions at different solid fractions in terms of dimensionless shear rate and shear stress (see text). The continuous line is the model fitted to data with $n = 0.36$. Inset: effective flow curves (filled and empty symbols correspond to different rotation velocities of the inner cylinder); no steady flow in the grey regions.

## 4. Conclusion

This approach predicts the transition from homogeneous flows to shear-banding as a function of simple physical parameters. In particular it suggests that the fundamental explanation of the transition is not in itself the repulsive vs attractive character change in the interaction type but rather the evolution of the characteristic time of restructuring associated with this change with usually studied systems. Due to its simplicity, this model, strictly devised for understanding this transition, obviously cannot predict the transient trends of the rheological behavior of these systems. However this model suggests basic ingredients to be taken into account in more sophisticated models. It also provides a simple, basic frame for comparing different systems, analyzing experimental data and devising materials with specific properties by tuning $D$ through one of the physical parameters it contains.

## References

[1] A.J. Liu, and S.R. Nagel, *Nature*, 396, 21 (1998)
[2] G. Ovarlez, S. Rodts, X. Chateau, P. Coussot, *Rheol. Acta*, 48, 831 (2009)

[3] F. Pignon, A. Magnin, and J.M. Piau, *J. Rheol.*, 40, 573 (1996); P. Coussot, Q.D. Nguyen, H.T. Huynh, and D. Bonn, *Phys. Rev. Lett.*, 88, 175501 (2002); P.C.F. Moller, S. Rodts, M.A.J. Michels, D. Bonn, *Phys. Rev. E* 77, 041507 (2008)

[4] L. Bécu, S. Manneville, and A. Colin, *Phys. Rev. Lett.*, 96, 138302 (2006); P.C.F. Moller et al., *Phil. Trans. R. Soc. A,* 367, 5139 (2009)

[5] P. Coussot et al., *J. Non-Newt. Fluid Mech.*, 158, 85 (2009)

[6] N. Roussel, R. Le Roy and P. Coussot, *J. Non-Newt. Fluid Mech.*, 117, 85 (2004)

[7] S.M. Fielding, M.E. Cates and P. Sollich, *Soft Matter*, 5, 2378 (2009)

[8] S. A. Rogers, D. Vlassopoulos, and P. T. Callaghan, *Phys. Rev. Lett.*, 100, 128304 (2008)

[9] P. Sollich, F. Lequeux, P. Hébraud, and M.E. Cates, *Phys. Rev. Lett.*, 78, 2020 (1997)

[10] J. Goyon et al., *Nature,* 454, 84 (2008) ; L. Bocquet, A. Colin, A. Ajdari, *Phys. Rev. Lett.*, 103, 036001 (2009)

[11] J. Vermant and M.J. Solomon, *J. Phys. Cond. Matter*, 17, 18 (2005)

[12] H.M. Princen and A.D. Kiss, *J. Colloid Interface Sci.*, 128, 176-187 (1988)

[13] S.A. Khan, C.A. Schnepper and C.A. Armstrong, *J. Rheol.*, 32, 69-92 (1998); A.D. Gopal, and D.J. Durian, *J. Colloid Interface Sci.*, 213, 169-178 (1999)

[14] R.I. Tanner, *Engineering rheology* (Clarendon Press, Oxford, 1985)

[15] P.D. Olmsted, *Rheol. Acta*, 47, 283 (2008)

[16] M. Cloitre, R. Borrega, F. Monti, and L. Leibler, *Phys. Rev. Lett.*, 90, 068303 (2003); S. Tcholakova, N. D. Denkov, K. Golemanov, K. P. Ananthapadmanabhan, and A. Lips, *Phys. Rev. E* 78, 011405 (2008)

[17] A.L. Biance, S.Cohen-Addad, R. Hohler, *Soft Matter*, 5, 4672 (2009)

[18] A. Ragouilliaux et al., *Phys. Rev. E*, 76, 051408 (2007)

[19] G. Ovarlez, R. Hohler, S. Cohen-Addad, Private communication (2010)

[20] S. Rodts, J.C. Baudez, and P. Coussot, *Europhysics Letters*, 69, 636-642 (2005)

[21] J.S. Raynaud et al., *J. Rheol.*, 46, 709 (2002)

[22] T. Divoux, D. Tamarii, C. Barentin, S. Manneville, *Phys. Rev. Lett.* 104, 208301 (2010)